\documentclass[aps,prd,twocolumn,superscriptaddress,longbibliography,nofootinbib]{revtex4-2}
\usepackage[utf8]{inputenc}
\usepackage{color}
\usepackage{graphicx}
\usepackage{subfigure}
\usepackage{xcolor}
\usepackage[normalem]{ulem}
\usepackage[pdftex,breaklinks,colorlinks,linkcolor=blue,citecolor=teal,anchorcolor=red,urlcolor=cyan]{hyperref}
\usepackage{orcidlink}
\usepackage{multirow}

\usepackage{amsmath,amssymb,amsfonts}

\def\prl{Phys. Rev. Lett.}
\def\prd{Phys. Rev. D}

\def\apj{Astrophys. J.}

\def\apjl{Astrophys. J. Lett.}

\def\pau_p{Prog. Theor. Phys.}

\def\mnras{Mon. Not. R. Astron. Soc.}

\def\physrep{Phys. Rep.}

\def\jcap{J. Cosmology Astropart. Phys}

\def\pasp{Publ. Astron. Soc. Pacific}
\def\mstar{\left( \frac{M_*}{M_\odot} \right)}
\def\rstar{\left( \frac{R_*}{R_\odot} \right)}
\def\mbh{\left( \frac{M_{\rm BH}}{M_\odot} \right)}
\def\vten{\left( \frac{v_\infty}{10^6 \,\mbox{cm/s}} \right)}

\begin{document}

\title{Can ``premature collapse" form black holes in the upper and lower mass gaps?}

\author{Thomas W.~Baumgarte\orcidlink{0000-0002-6316-602X}}
\email{tbaumgar@bowdoin.edu}
\affiliation{Department of Physics and Astronomy, Bowdoin College, Brunswick, Maine 04011, USA}

\author{Stuart L.~Shapiro\orcidlink{0000-0002-3263-7386}}
\email{slshapir@illinois.edu}
\affiliation{Department of Physics, University of Illinois at Urbana-Champaign, Urbana, Illinois 61801}
\affiliation{Department of Astronomy and NCSA, University of Illinois at Urbana-Champaign, Urbana, Illinois 61801}

\begin{abstract}
Observations of gravitational waves from binary black hole mergers, including the recent signals GW231123 and GW230529, have revealed multiple progenitor black holes in the so-called upper and lower mass gaps, respectively.  It is generally assumed that massive stars cannot form black holes in the upper mass gap because pair instabilities in the late stage of stellar evolution disrupt the stars, whereas the lower mass gap refers to the gap between the maximum allowed neutron star mass and the smallest black hole mass expected to form in supernova explosions.   Here we explore a ``premature collapse" scenario in which upper mass gap stars collapse and form black holes {\em before} they reach the late stage of stellar evolution. The mechanism for triggering a premature collapse is the capture of a smaller black hole, possibly primordial in nature.  A similar capture scenario can occur to produce black holes in the lower mass gap.  At least for massive stars, typical stellar rotation rates would likely result in rapidly rotating black holes in such a scenario, naturally explaining the rapid spins inferred from GW231123.    Even though our estimates hinge on several parameters with rather large uncertainties, they suggest that, at least in galactic disks, the likelihood of such a capture is small for stars in the upper mass gap, but may lead to a significant population of black holes in the lower mass gap and, in fact, even below the lower mass gap.
\end{abstract}

\maketitle

Since the spectacular first direct detection of gravitational waves in 2015 \cite{LIGO_GW150914}, multiple observed signals are believed to originate from binary black hole mergers with progenitor black holes in either the ``upper mass gap", between around 60 and $130 M_\odot$, or the ``lower mass gap", between about 3 and $5 M_\odot$.

Stellar evolution theory predicts that presupernova stars with core masses between about 60 and $130 M_\odot$ explode completely in a pair-instability supernova, leaving behind no compact remnant (see, e.g., \cite{WooH21} and references therein).  While the exact mass limits are subject to a number of different assumptions (see, e.g., \cite{CrooSG25,GotMILRI25,PopdeM25} for recent discussions), this prediction gives rise to the notion of the {\em upper mass gap} in which no black holes are expected to form from stellar evolution and supernova collapse.  And yet, observations of gravitational wave signals have revealed several progenitor black holes with masses in this range.  A particularly interesting example is the recently observed signal GW231123 \cite{LIGO_GW231123}, which is consistent with the merger of two black holes with masses $137^{+22}_{-17} M_\odot$ and $103^{+30}_{-52} M_\odot$.  Moreover, both black holes are believed to be rapidly rotating, with (dimensionless) spins $\chi \equiv c J / (G M^2)$ of $0.9^{+0.10}_{-0.19}$ and $0.80^{+0.20}_{-0.51}$ respectively. 

Evidently, then, some process other than the supernova explosion of a massive star must have resulted in the upper mass gap black hole (or holes) observed in GW231123 and other gravitational wave signals.  Perhaps the most promising such process is the hierarchical formation of massive black holes from the successive mergers of less massive black holes (see, e.g, \cite{SteOM25,LiTXF25,DelRCNMFPMO25,LiF25} for analyses in the context of GW231123), even though such mergers may also lead to large recoil speeds, ejecting the remnant from clusters or even its host galaxy, and may not lead to rapidly spinning remnants \cite{KirKGR25} (but see \cite{BamSRT25} for a fully relativistic simulation of a compact cluster of arbitrary mass black holes that produced a bound, third-generation black hole merger remnant with spin $\chi \simeq 0.8-0.9$; see also the discussion in \cite{LIGO_GW231123}). Some authors have also considered alternative formation scenarios, including primordial black holes (PBHs) \cite{CarCGHK24,YuaCL25,deLucFR25} and the collapse of Population III stars \cite{TanLFW25}.\footnote{See also \cite{CucBCS25} for an exploration of whether GW231123 might have been emitted from cusps or kinks in cosmic strings.}

The {\em lower mass gap} refers to the range between the maximum allowed mass of neutron stars, about $3 M_\odot$, and the smallest black hole expected to form in supernova explosions, around $5 M_\odot$.  The exact limits are again subject to a number of uncertainties, and even the existence of this mass gap has been debated by some authors (e.g.~\cite{RayFK25}).  As for the upper mass gap, several gravitational wave detections hint at progenitor black holes with masses in the lower mass gap, including the recently observed signal GW230529, for which the primary is estimated to have a mass of $3.6^{+0.8}_{-1.2} M_\odot$ and a spin parameter of $\chi = 0.44^{+0.40}_{-0.37}$ (see \cite{LIGO_GW230529}, see also \cite{FisBWS25} for Gaia observations of lower mass gap objects).  Several authors have discussed different evolutionary pathways to forming such black holes (e.g.~\cite{AfrM25}), including low-mass black hole formation in supernovae (e.g.~\cite{FryBWDKH12,BocF24,ZhuHKZTSQ24,Xinetal24,BurWV25}), the merger of two neutron stars (e.g.~\cite{MahCGAFSA25,DhaCRKSLP25}, but see also \cite{MarPBD25}), and PBHs \cite{CarCGHK24,HuaYCL24}. Alternatively, at least some observed companions in the lower mass gap, such as GW190814 \cite{LIGO_GW190814} at $2.59^{+0.08}_{-0.09} M_\odot$, may actually be massive neutron stars supported by a stiff equation of state and/or rotation (see \cite{TsoRS20}). 

In this Letter we explore yet another alternative formation scenario for black holes possibly in both mass gaps: the collapse of stars with masses in the respective mass gaps, whose collapse is induced by the capture of a small black hole.  For upper mass gap stars, the collapse is induced {\em before} the late stellar evolution and subsequent stellar disruption.  In the following we refer to this scenario as a ``premature collapse" for both mass gaps.

We start by noting that such a collapse would quite naturally lead to rapidly spinning black hole remnants, at least for massive progenitor stars.  While we are unaware of fully relativistic numerical simulations of the collapse of rotating main-sequence stars to black holes, other somewhat similar simulations may provide some guidance.  For example, simulations of the collapse of rapidly rotating supermassive stars dominated by radiation pressure show that most of the mass forms a rapidly rotating black hole with spin parameter $\chi \simeq 0.7$ (see, e.g., \cite{ShiS02,ShaS02,ShiSUU16}).  Similarly, simulations of rotating neutron stars harboring a small black hole at their centers suggest that most of the mass is accreted, and that the black hole's final value of the spin parameter $\chi$ is therefore close to the progenitor star's initial value of $\chi$ (see \cite{EasL19}).  

For massive main-sequence stars, $\chi$ can be estimated from the empirical relations between average angular momenta and stellar mass provided by, for example, Kraft \cite{Kra70} and Kawaler \cite{Kaw87}.   Adopting Eq.~(3) of \cite{Kaw87} we have
\begin{equation} \label{kraft}
\langle J_{\rm MS} \rangle = (8.95 \pm 0.55) \times 10^{49} \mbox{g\,cm}^2 \mbox{s}^{-1} \left(\frac{M}{M_\odot} \right)^{2.09 \pm 0.05}
\end{equation}
for early-type, i.e.~massive stars with masses greater than about $1.5 M_\odot$.  Because the exponent in (\ref{kraft}) is close to 2, the dimensionless quantity $\langle \chi_{\rm MS} \rangle  = c \langle J_{\rm MS} \rangle /(GM^2)$ is nearly independent of the stellar mass and is well approximated by
\begin{equation} \label{chi_MS}
\langle \chi_{\rm MS} \rangle \simeq 10.
\end{equation}
Similar values can also be computed from the data provided, for example, in \cite{McNal65}. Later-type, i.e.~less massive stars, carry significantly less angular momentum (see, e.g., the discussion in \cite{Tas00}); for the Sun, for example, $\chi_\odot \simeq 0.2$. More recent observations (e.g.~\cite{HuaGM10} and references therein) suggest that stellar rotation rates show significant variation and are affected by several factors, including environment and age.  Clearly, a star exceeding the Kerr upper limit $\chi_{\rm max} = 1$ allowed for black holes cannot collapse without losing some angular momentum.  Invoking the numerical simulations mentioned above, however, together with the large values of $\chi$ for many massive stars, we speculate that the black hole formed in the premature collapse of such a star will be rapidly rotating, with $\chi \lesssim 1$, in accordance with the observations of GW231123.  


A possible scenario for triggering a premature collapse is the capture of a smaller black hole with mass $M_{\rm BH}$.  As long as $M_{\rm BH} \ll M_*$, where $M_*$ is the stellar mass, the black hole is unlikely to disrupt the star tidally before being swallowed.  Once inside the star, such an ``endoparasitic" black hole accretes the stellar material, ultimately triggering stellar collapse (see, e.g., \cite{CapPT13,AbrBW18,EasL19,GenST20,RicBS21,BauS24a,BauS24b} in the context of neutron stars and \cite{BauS25a} for massive stars).  While the accretion process may take {\em many} stellar dynamical timescales (see Eqs.~(\ref{tau_acc}) and (\ref{tau_dyn}) below), the process ends with a dynamical collapse of the entire star.  We note that we {\em assume} here that the capture of a black-hole intruder during the main-sequence phase of an upper mass gap star would lead to stellar consumption by the black hole rather than explosive disruption, but suggest that the exact dynamical process should be explored in future numerical simulations.

In order to explore the viability of such a scenario we model the star as a $\Gamma = 4/3$ ($n = 3$) polytrope in the following, i.e.~we approximate its equation of state as
\begin{equation} \label{EOS}
P = K \rho^\Gamma,~~~~~~\Gamma = 4/3.
\end{equation}
This approximation becomes exact in the limit that the star is dominated by radiation pressure, as in the upper mass gap; the same approximation serves as the standard Eddington model for the lower mass gap.  In (\ref{EOS}) $K$ is related to the stellar mass $M_*$ by
\begin{equation}
K = \frac{\pi^{1/3} G}{4^{2/3} (\xi_3^2 |\theta'(\xi_3)|)^{2/3}} M_*^{2/3},
\end{equation}
where $\xi_3^2 |\theta'(\xi_3)| \simeq 2.018$ for $n = 3$ (see, e.g., \cite{ShaT83} for a textbook treatment).

We first estimate the accretion radius $r_a = GM_{\rm BH}/a^2$ for a black hole of mass $M_{\rm BH}$ at the stellar center, where  
\begin{equation}
a^2 = \Gamma K \rho^{\Gamma - 1}
\end{equation}
is the sound speed.   Combining the above expressions and evaluating them at the stellar center we obtain
\begin{equation} \label{ra}
r_a \simeq 3^{2/3} (\xi_3^2 |\theta'(\xi_3)|)^{2/3} \delta^{-1/3}
\left(\frac{M_{\rm BH}}{M_*}\right) R_*,
\end{equation}
where $\delta = \rho_c / \bar \rho$ with $\bar \rho = 3 M /(4 \pi R^3)$ is the central condensation and takes the value of $\delta \simeq 54.18$ for $n = 3$.  Evidently, as long as $M_{\rm BH} \ll M_*$ the accretion radius is much smaller than the stellar radius.  In this case we may model the black-hole mass accretion rate as Bondi accretion at the stellar center,
\begin{equation} \label{bondi1}
\dot M_{\rm BH} = \frac{4 \pi \lambda G^2 M_{\rm BH}^2 \rho_c}{a_c^3}
\end{equation}
(see \cite{Bon52} as well as \cite{ShaT83} for a textbook treatment).\footnote{Eq.~(\ref{ra}) implies $G M_{\rm BH}/c^2 \ll r_a \ll R_*$, so that for a uniformly rotating star the flow will be approximately spherical at $r_a$ (i.e.~Bondi accretion), but will form an accretion disk closer to the black hole.}  
Using the above expressions for $a$ and $K$, as well as $\lambda = 2^{-1/2}$ for $n = 3$, (\ref{bondi1}) becomes
\begin{equation}  \label{bondi2}
\dot M_{\rm BH} = A \left( \frac{M_*}{M_\odot} \right)^{-1/2} 
\left( \frac{R_*}{R_\odot} \right)^{-3/2} M_{\rm BH}^2, 
\end{equation}
where $M_\odot$ and $R_\odot$ are the solar mass and radius, respectively, and where we have abbreviated
\begin{equation}
A \equiv 9 \lambda \xi_3^2 | \theta(\xi_3) | \delta^{1/2} \frac{G^{1/2}}{M_\odot^{1/2} R_\odot^{3/2}} = 2.97 \times 10^{-35} \mbox{g}^{-1} \mbox{s}^{-1}.
\end{equation}
Integrating $M_{\rm BH}$ in (\ref{bondi2}) over time from some initial mass $M_{\rm BH}^{\rm init}$, assumed to be much less than $M_*$, to its final mass $M_{\rm BH}^{\rm fin} \simeq M_*$ we obtain the accretion time scale\footnote{In carrying out the integration we keep $M_*$ and $R_*$ constant.  This is an excellent approximation while $M_{\rm BH} \ll M_*$, which dominates the accretion timescale.} 
\begin{align} \label{tau_acc}
\tau_{\rm acc} & \simeq \frac{1}{A} \left( \frac{M_*}{M_\odot} \right)^{1/2} \left( \frac{R_*}{R_\odot} \right)^{3/2} \frac{1}{M_{\rm BH}^{\rm init}} \nonumber \\
& \simeq 16.8 \mbox{\,s} \left( \frac{M_*}{M_\odot} \right)^{1/2} \left( \frac{R_*}{R_\odot} \right)^{3/2} \left( \frac{M_{\rm BH}^{\rm init}}{M_\odot} \right)^{-1}.
\end{align}
For most scenarios discussed below, $\tau_{\rm acc}$ is longer than or comparable to the dynamical timescale
\begin{equation} \label{tau_dyn}
\tau_{\rm dyn} \simeq \left( \frac{R^3}{GM} \right)^{1/2} 
\simeq 1.6 \times 10^3 \mbox{\,s} \left( \frac{M_*}{M_\odot} \right)^{-1/2} \left( \frac{R_*}{R_\odot} \right)^{3/2}
\end{equation}
and hence dominates the timescale for triggering a premature collapse.  Since either timescale is significantly shorter than the stellar lifetime for the black-hole masses considered below, premature collapse triggered by black hole capture is viable at least vis-\`a-vis timescales.

In order for a black hole to be captured by a star, it has to lose a sufficient amount of energy during its first passage through the star. When the two objects are at a large separation and move with a relative speed $v_\infty$, the total energy of the system is the kinetic energy $K_\infty \simeq M_{\rm BH} v_\infty^2/2$ (where we assume $M_{\rm BH} \ll M_*$).  If hydrodynamical friction during the first passage dissipates an energy greater than $K_\infty$, then the black hole may still reemerge from the star, but it can no longer escape from the star's gravitational potential.  Instead, it will return to the star for subsequent passages, ultimately being swallowed by the star and settling down toward the stellar center (see, e.g., Fig.~5 in \cite{AbrBW18} and Fig.~1 in \cite{BauS24a} for illustrations).  Equating $K_\infty$ with the energy dissipated during the first collision leads to the estimate
\begin{equation} \label{min_mass}
M_{\rm BH}^{\rm min} \simeq \frac{1}{3 \ln \Lambda}  \left( \frac{v_\infty}{v_{\rm esc}} \right)^2 M_*
\end{equation}
for the minimum black-hole mass for capture (see, e.g., Eq.~(11) in \cite{BauS24a}).  Here 
\begin{align} \label{vesc}
v_{\rm esc} & = \left( \frac{2 G M_*}{R_*} \right)^{1/2} \nonumber \\ 
& \simeq
6.19 \times 10^7 \mbox{cm\,s}^{-1} 
\mstar^{1/2} \rstar^{-1/2}
\end{align}
is the escape speed from the surface of the star and $\ln \Lambda$ the Coulomb logarithm.  The relative speed $v_\infty$ at infinite separation depends on the stellar environment.  For the Galactic disk, for example, typical dispersion speeds $v_\infty$ are on the order of about $10 - 30$ km/s.  Using these values we have
\begin{equation} \label{min_mass2}
M_{\rm BH}^{\rm min} \simeq 8.70 \times 10^{-6} M_\odot  
\rstar \vten^2,
\end{equation}
where we have approximated $\ln \Lambda \simeq 10$.
We caution that the timescale for capture is typically longer than that for accretion estimated in (\ref{tau_acc}), because it may take the black hole multiple passages through the star before remaining inside.  The exact number of passages, and the time spent outside the star between them, depends sensitively on both $v_\infty$ and $M_{\rm BH}$ as well as the stellar properties (see, e.g., \cite{AbrBW18,BauS24a}).  In the following we assume that the resulting capture timescale is shorter than the stellar lifetime for black holes above $M_{\rm BH}^{\rm min}$, which is the case for sufficiently small $v_\infty$.

We next tackle the question of the origin of the captured black hole.  One option might be supernova collapse of another star, with a mass below the mass gap.  Assuming that all stars in close proximity have approximately the same age, however, such a smaller-mass star would evolve more slowly than the more massive stars.  That means that smaller-mass black holes have not yet been formed by the time the more mass star would be disrupted by pair instabilities.

Another option for the origin of the smaller captured black hole are primordial black holes.  The most promising mass window for PBHs as the source of the dark matter content of the Universe extends from about $10^{-16} M_\odot$ to about $10^{-10} M_\odot$ (see, e.g., Fig.~10 in \cite{CarKSY21} and Fig.~1 in \cite{CarK20}), but from our estimate (\ref{min_mass2}) we see that such black holes are unlikely to be captured by main-sequence stars.  However, PBHs in two other windows around $10^{-6} M_\odot$ and $1 M_\odot$ may constitute a fraction $f_{\rm PBH}$ up to about 10\% of the dark matter content.  According to Eq.~(\ref{min_mass2}), stars in the lower mass gap might be able to capture PBHs in both of these windows.  However, stars in the upper mass gap, with their larger stellar radii $R_*$, would likely be unable to capture PBHs as small as $10^{-6} M_\odot$.

The rate at which a single star collides with a hypothetical PBH can be estimated from
\begin{equation} \label{coll1}
\dot {\mathcal N} = \sigma v_\infty n,
\end{equation}
where $\sigma$ is the cross-section for a collision and $n$ the PBH number density.  The former is
\begin{equation} \label{sigma}
\sigma = \pi R_*^2 \left(1 + \left(\frac{v_{\rm esc}}{v_\infty} \right)^2 \right)
\end{equation}
(see, e.g, \cite{AbrBBGJQ09}) where, in the following, we assume that the second term dominates over the first.  We also write
\begin{equation} \label{n}
n = f_{\rm PBH} \frac{\rho_{\rm DM}}{M_{\rm BH}} 
= f_{\rm PBH} \frac{\rho_{\rm DM}}{\rho_{\rm DM}^{\rm loc}} \frac{10^{-24} \mbox{\,g\,cm}^{-3}}{M_{\rm BH}}.
\end{equation}
Here we have adopted $\rho_{\rm DM}^{\rm loc} \simeq 10^{-24} \mbox{\,g\,cm}^{-3}$ for the local dark-matter density in the solar neighborhood \cite{McKPH15}, as the fiducial environment we will adopt to illustrate the process is the Galactic disk.  The actual dark-matter density might be significantly higher than $\rho_{\rm DM}^{\rm loc}$ in dense regions of stars, the cores of active galactic nuclei or ultra-faint dwarf galaxies (see, e.g., \cite{SmiGLM24}, who adopt a generalized Navarro-Frenk-White profile to model the Galactic dark-matter density and estimate collision rates between PBHs and white dwarfs).  We also note, however, that the dark-matter density in globular clusters is either very low or possibly zero, so by contrast with hierarchical merger scenarios, these presumably would not provide a suitable environment for this scenario.

\begin{widetext}
Inserting (\ref{vesc}), (\ref{n}), and (\ref{sigma}) into (\ref{coll1}) we now obtain
\begin{align} \label{Ndot}
\dot {\mathcal N} & \simeq 2 \pi G R_* \frac{10^{-24} \mbox{\,g\,cm}^{-3}}{v_\infty} \frac{M_*}{M_{\rm BH}} f_{\rm PBH} \frac{\rho_{\rm DM}}{\rho_{\rm DM}^{\rm loc}} 
\simeq 2.9 \times 10^{-26} \mbox{s}^{-1} \rstar \vten^{-1} \left( \frac{M_*}{M_{\rm BH}} \right) f_{\rm PBH} \frac{\rho_{\rm DM}}{\rho_{\rm DM}^{\rm loc}}.
\end{align}
 In order to find the likelihood that a star collides with a PBH while on the main sequence, we multiply the rate (\ref{Ndot}) with the main-sequence lifetime $\tau_{\rm MS}$.  Approximating 
\begin{equation}
\tau_{\rm MS} \simeq 3 \times 10^{17} \mbox{s}~\mstar^{-\alpha},
\end{equation}
where $\alpha$ is between about 2 and 3, we obtain
\begin{align} \label{N}
{\mathcal N}_{\rm tot} & \simeq \dot {\mathcal N} \tau_{\rm MS} 
\simeq 8.8 \times 10^{-9 }~\mstar^{1-\alpha} \rstar
\vten^{-1} \mbh^{-1}
f_{\rm PBH} \frac{\rho_{\rm DM}}{\rho^{\rm loc}_{\rm DM}}
\end{align}
for the total number of collisions with PBHs during a stellar lifetime. Even though the collision {\em rate} (\ref{Ndot}) between stars and PBHs increases with increasing $M_*$, the total {\em number} of collisions (\ref{N}) during the stellar main-sequence lifetime increases with decreasing $M_*$ due to the longer lifetime of less massive stars.
\end{widetext}

We adopt a simple analytic initial mass function (IMF)
\begin{equation}
\psi(M_*) d\left(\frac{M_*}{M_\odot}\right) = \psi_0 \left( \frac{M_*}{M_\odot} \right)^{-2.35} d\left(\frac{M_*}{M_\odot}\right),
\end{equation} 
where $\psi_0 = 2 \times 10^{-12} \mbox{\,stars\,pc}^{-3} \mbox{\,yr}^{-1}$ 
(Salpeter IMF; \cite{Sal55}) to crudely estimate the total number of black holes that might have formed from premature collapse induced by capture of primordial black holes.  Focusing on the upper mass gap first, the Salpeter IMF predicts that about a fraction $10^{-4}$ of stars have masses between about 60 and $130 M_\odot$.  Assuming there are $\sim 10^{11}$ stars in the Galactic disk, this suggests that about $10^7$ of these stars are in the upper mass gap (which may well be an overestimate).  Adopting $M_* = 100 M_\odot$ and $R_* = 100 R_\odot$ as well as $\alpha = 2.5$ as approximate values in (\ref{N}), we then expect about 
\begin{equation}
N_{\rm BH}^{\rm upper} \simeq 10^{-2} \vten^{-1}
\mbh^{-1}
f_{\rm PBH} \frac{\rho_{\rm DM}}{\rho^{\rm loc}_{\rm DM}}
\end{equation}
black holes in the upper mass gap in our Galaxy.  This estimate still depends, of course, on assumptions on the most realistic or favorable PBH masses and galactic environments.  Adopting the PBH mass window $M_{\rm BH} \simeq 1 M_{\odot}$ with $f_{\rm PBH} \simeq 0.1$, together with $\rho_{\rm DM}/\rho_{\rm DM}^{\rm loc} \simeq 1$ and $v_\infty = 10^6 \mbox{cm\,s}^{-1}$, we find $N_{\rm BH}^{\rm upper} \simeq 10^{-3}$.   For comparison, the Salpeter IMF yields about $10^8$ stars between 10 and $60 M_\odot$ which likely undergo supernova collapse to black holes.   The small value for $N_{\rm BH}^{\rm upper}$ in comparison therefore suggests that it is highly unlikely that an upper mass gap progenitor in a binary black hole merger was formed by premature collapse induced by PBH capture, even if $\rho_{\rm DM}/\rho_{\rm DM}^{\rm loc}$ were significantly larger than unity. 

We similarly estimate from the Salpeter IMF that the fraction of stars in the lower mass gap, between about 3 and $5 M_\odot$, is about 0.4\%, meaning that there should be about $4\times 10^{8}$ such stars in our Galaxy.  Now adopting $M_* = 4 M_\odot$ and $R_* = 3 R_\odot$ in (\ref{N}) we expect
\begin{equation}
N_{\rm BH}^{\rm lower} \simeq 1.3 \vten^{-1} \mbh^{-1}
f_{\rm PBH} \frac{\rho_{\rm DM}}{\rho^{\rm loc}_{\rm DM}}
\end{equation}
black holes in the lower mass gap in our Galaxy.  Adopting $M_{\rm BH} \simeq 10^{-6} M_\odot$ (assuming that black holes with these masses can be captured by these stars; see the discussion above), together with $f_{\rm PBH} \simeq 0.1$, and using $\rho_{\rm DM}/\rho_{\rm DM}^{\rm loc} \simeq 1$  and $v_\infty = 10^6 \mbox{cm\,s}^{-1}$ as before, we now estimate $N_{\rm BH}^{\rm lower} \simeq 10^5$.  Comparing this with the $\sim 10^8$ Galactic disk black holes believed to have formed in supernova explosions, we see that a small but non-negligible fraction of all black holes might indeed be in the lower mass gap.  The exact number depends, of course, on $v_\infty$, $f_{\rm PBH}$, and $\rho_{\rm DM}$, as well as the very existence of PBHs.  As we argued above, $\rho_{\rm DM}$ may be significantly larger than $\rho_{\rm DM}^{\rm loc}$, which would increase the number of lower-mass gap black holes accordingly.

As an aside we note that, according to the same estimates based on the Salpeter IMF, about 2\% of main sequence stars should be in the {\em neutron-star mass range} between 1 and $2 M_\odot$.  Eq.~(\ref{N}) then predicts about 
\begin{equation} \label{N_ns}
N_{\rm BH}^{\rm NS} \simeq 17 \vten^{-1} \mbh^{-1}
f_{\rm PBH} \frac{\rho_{\rm DM}}{\rho^{\rm loc}_{\rm DM}}
\end{equation}
collisions of these main sequence stars with PBHs would lead to the formation of black holes in the neutron-star mass range.  Adopting $M_{\rm BH} = 10^{-6} M_\odot$ with $f_{\rm PBH} \simeq 0.1$ again, as well as $\rho_{\rm DM}/\rho_{\rm DM}^{\rm loc} \simeq 1$  and $v_\infty = 10^6 \,\mbox{cm\,s}^{-1}$, the estimate (\ref{N_ns}) yields $N_{\rm BH}^{M_\odot} \simeq 10^6$.  From the Salpeter IMF we also estimate that there are about $4 \times 10^8$ neutron stars in the Galaxy that we assume arise from the collapse of stars between $4 M_\odot$ and $10 M_\odot$ at the endpoint of their evolution (even though some authors adopt a number close to $10^9$).  This suggests again that a small but significant fraction of compact objects in the neutron-star mass range might, in fact, be black holes that are formed from low-mass stars that capture a PBH.  Again, a larger value of $\rho_{\rm DM}$ would result in a larger value of this fraction.  Moreover, this scenario would lead to black holes with masses {\it below} the putative lower mass gap!

A similar scenario leading to the formation of black holes in the neutron-star mass range is the capture of PBHs by neutron stars (see, e.g., \cite{CapPT13,AbrBW18,EasL19,GenST20,BauS24a}).  Because of the high escape speed from the surface of neutron stars, they are able to capture significantly smaller PBHs than main-sequence stars (see the estimate (\ref{min_mass}) above).

We again caution that several terms in the above estimates are not well known and subject to significant debate -- including, of course, the very existence of PBHs themselves.  If they exist, however, some are likely to collide with stars and may induce them to collapse and form black holes in the upper and lower mass gaps.  While, in the upper mass gap, such a premature-collapse scenario would likely lead to rapidly-rotating black holes -- as observed, for example, in GW231123 -- the event rates appear to be small enough to make this scenario highly unlikely.  On the other hand, our crude estimates suggest that black-hole capture inducing collapse may lead to a significant fraction of all Galactic disk black holes in the lower mass gap, and hence may explain progenitors like the primary in GW230529.  Moreover, our estimates suggest the possible existence of a population of black holes in the neutron star mass range, so that a significant fraction of compact objects between about 1 and $2 M_\odot$ might be black holes and thus below the putative lower mass gap.

\acknowledgments

We would like to thank the anonymous referees for helpful comments.  This work was supported in parts by National Science Foundation (NSF) grant PHY-2308821 to Bowdoin College, as well as NSF grant PHY-2308242 to the University of Illinois at Urbana-Champaign.

%

\end{document}